\documentclass[12pt]{article}
\usepackage{graphicx,amssymb,amsfonts,amsmath,epsfig,color}
\usepackage{mathrsfs}
\usepackage{cite}
\newcounter{mnotecount}

\newcommand{\mnotex}[1]
{\protect{\stepcounter{mnotecount}}$^{\mbox{\footnotesize $\bullet$\themnotecount}}$
\marginpar{
\raggedright\tiny\em
$\!\!\!\!\!\!\,\bullet$\themnotecount: #1} }

\makeatletter
\DeclareSymbolFont{AMSb}{U}{msb}{m}{n}
\DeclareSymbolFontAlphabet{\mathbb}{AMSb}
\renewcommand{\section}{\@startsection{section}{1}{\z@}%
                                    {-7ex \@plus -1ex \@minus -.2ex}%
                                    {2.5ex \@plus.2ex}%
                                    {\normalfont\large\scshape\centering}}
\renewcommand{\subsection}{\@startsection{subsection}{2}{\z@}%
                                       {-5ex \@plus -1ex \@minus -.2ex}%
                                       {1.5ex \@plus.2ex}%
                                       {\normalfont\normalsize\scshape}}
\renewcommand{\subsubsection}{\@startsection{subsubsection}{3}{\z@}%
                                       {-5ex \@plus -1ex \@minus -.2ex}%
                                       {1.5ex \@plus.2ex}%
                                       {\normalfont\normalsize\scshape}}

\renewcommand\@seccntformat[1]{\ignorespaces\csname #1name\endcsname\space
                               \csname the#1\endcsname.\quad}   
\newdimen\captionmargin
\newdimen\captionindent
\newdimen\captionwidth
\newcommand{\captionfont}{\slshape}
\newcommand\@captionlabel[1]{\textsc{#1:}\space}
\long\def\@makecaption#1#2{%
  \vskip\abovecaptionskip
  \captionwidth\hsize
  \advance\captionwidth -2\captionmargin
  \sbox\@tempboxa{\@captionlabel{#1}\captionfont #2}%
  \ifdim \wd\@tempboxa >\captionwidth
    \ifdim\captionindent>\z@
      \advance\captionwidth -\captionindent
      \hskip\captionindent
    \fi
    \hskip\captionmargin
    \parbox[t]{\captionwidth}{\leavevmode\hskip-\captionindent
      \@captionlabel{#1}\captionfont #2}%
  \else
    \global \@minipagefalse
    \hb@xt@\hsize{\hfil\box\@tempboxa\hfil}%
  \fi
  \vskip\belowcaptionskip}
\def\eqnarray{%
   \stepcounter{equation}%
   \def\@currentlabel{\p@equation\theequation}%
   \global\@eqnswtrue
   \m@th
   \global\@eqcnt\z@
   \tabskip\@centering
   \let\\\@eqncr
   $$\everycr{}\halign to\displaywidth\bgroup
       \hskip\@centering$\displaystyle\tabskip\z@skip{##}$\@eqnsel
      &\global\@eqcnt\@ne$\;\hfil{##}$\hfil
      &\global\@eqcnt\tw@$\;\displaystyle{##}$\hfil\tabskip\@centering
      &\global\@eqcnt\thr@@ \hb@xt@\z@\bgroup\hss##\egroup
         \tabskip\z@skip
      \cr}
\ifcase \@ptsize
\or
\or
\fi
  \divide\@tempdima\baselineskip
  \@tempcnta=\@tempdima
\makeatother
\begin{document}

\renewcommand{\theequation}{\arabic{section}.\arabic{equation}}
\renewcommand{\thefigure}{\arabic{figure}}
\newcommand{\gapprox}{%
\mathrel{%
\setbox0=\hbox{$>$}\raise0.6ex\copy0\kern-\wd0\lower0.65ex\hbox{$\sim$}}}
\textwidth 165mm \textheight 220mm \topmargin 0pt \oddsidemargin 2mm
\def\ib{{\bar \imath}}
\def\jb{{\bar \jmath}}

\newcommand{\ft}[2]{{\textstyle\frac{#1}{#2}}}
\newcommand{\be}{\begin{equation}}
\newcommand{\ee}{\end{equation}}
\newcommand{\bea}{\begin{eqnarray}}
\newcommand{\eea}{\end{eqnarray}}
\newcommand{\Identity}{{1\!\rm l}}
\newcommand{\cx}{\overset{\circ}{x}_2}
\def\CN{$\mathcal{N}$}
\def\CH{$\mathcal{H}$}
\def\hg{\hat{g}}
\newcommand{\bref}[1]{(\ref{#1})}
\def\espai{\;\;\;\;\;\;}
\def\zespai{\;\;\;\;}
\def\avall{\vspace{0.5cm}}
\newtheorem{theorem}{Theorem}
\newtheorem{acknowledgement}{Acknowledgment}
\newtheorem{algorithm}{Algorithm}
\newtheorem{axiom}{Axiom}
\newtheorem{case}{Case}
\newtheorem{claim}{Claim}
\newtheorem{conclusion}{Conclusion}
\newtheorem{condition}{Condition}
\newtheorem{conjecture}{Conjecture}
\newtheorem{corollary}{Corollary}
\newtheorem{criterion}{Criterion}
\newtheorem{defi}{Definition}
\newtheorem{example}{Example}
\newtheorem{exercise}{Exercise}
\newtheorem{lemma}{Lemma}
\newtheorem{notation}{Notation}
\newtheorem{problem}{Problem}
\newtheorem{prop}{Proposition}
\newtheorem{rem}{{\it Remark}}
\newtheorem{solution}{Solution}
\newtheorem{summary}{Summary}
\numberwithin{equation}{section}
\newenvironment{pf}[1][Proof]{\noindent{\it {#1.}} }{\ \rule{0.5em}{0.5em}}
\newenvironment{ex}[1][Example]{\noindent{\it {#1.}}}

\thispagestyle{empty}


\begin{center}

{\LARGE\scshape Non-Singular Black Holes, the Cosmological Constant and Asymptotic Safety   \par}
\vskip15mm

\textsc{Ram\'{o}n Torres\footnote{E-mail: ramon.torres-herrera@upc.edu}}
\par\bigskip
{\em
Dept. de F\'{i}sica, Universitat Polit\`{e}cnica de Catalunya, Barcelona, Spain.}\\[.1cm]
%


\end{center}

\begin{abstract}
Quantum gravitational effects in black hole spacetimes with a cosmological constant $\Lambda$ are considered. The effective quantum spacetimes for the black holes are constructed by taking into account the renormalization group improvement of classical solutions obtained in the framework of Unimodular Gravity (a theory which is identical to General Relativity at a classical level). This allows us to avoid the usual divergences associated with the presence of a \textit{running} $\Lambda$.
The horizons and causal structure of the improved black holes are discussed taking into account the current observational bounds for the cosmological constant. It is shown that the resulting effective quantum black hole spacetimes are always devoid of singularities.
\end{abstract}




\setcounter{equation}{0}

\section{Introduction}

The pioneering works on gravitational collapse in the framework of General Relativity (GR) (see, for instance, \cite{O&S}\cite{LSM1965}) seemed to show that, under the appropriate circumstances, the formation of black holes
was unavoidable. Moreover, black holes appeared to be accompanied of an inner singularity, something that was later backed up with the development of the \textit{singularity theorems}\cite{Seno}. However, the formation of singularities (where GR cannot longer be used) has been considered by many as a weakness of the theory rather than as a real physical prediction.
In fact, it is usually expected that the inclusion of quantum theory in the description of black holes could avoid the existence of their singularities.
Indeed, some paradigms and heuristic models of \emph{non-singular} Black Holes inspired in different approaches to Quantum Gravity have appeared in the recent literature (see, for example, 
\cite{B&R,A&B2005,Hay2006,Frolov2014,dust2014,G&P2014,H&R2014} and references therein).

One of the most promising approaches to a consistent and predictive quantum theory of the gravitational field is Asymptotic Safety (AS) which was proposed by Steven Weinberg in 1976 \cite{Wein}. AS requires the existence of a non-trivial fixed point of the theory's renormalization group which controls the behaviour of the coupling constants. In this way, physical quantities could be safe from divergences in the ultraviolet regime, without being perturbatively renormalizable.
In the nineties, the advent of new functional renormalization group methods made possible the construction of an effective average action $\Gamma_k$ (that depends on the energy scale $k$ under consideration) and the associated flow equation for the gravitational field \cite{Reuter}. Since then, a variety of nonperturbative computations has been carried out
in this framework
\cite{Reuter,Litim,Percacci,Saueressig}
now known as Quantum Einstein Gravity (QEG) (see, for example, \cite{R&S} for a review).

Among the non-perturbative computations within the QEG framework let us mention those using the \textit{Einstein-Hilbert truncation} in which one takes into account only two couplings: Newton's \textit{constant} $G(k)$ and the cosmological \textit{constant} $\Lambda(k)$.
If one considers a four dimensional spacetime and momentarily puts aside $\Lambda$ one gets \cite{Reuter} a flow equation for the dimensionless Newton constant $g(k)$ [$\equiv k^2 G(k)$] from which one finds a non-trivial fixed point 
$g^* \in \Re^+$. In this way, as the energy scale grows, one gets that the gravitational coupling weakens, i.e., $G(k)\rightarrow 0$ (what can be justified as an antiscreening effect produced by the gravitons \cite{N&R}). It would seem that a natural consequence of this could be
the avoidance of the classical singularity in the interior of black holes \cite{B&R}\footnote{In fact, this true even if the number of dimensions is bigger than 4 \cite{B&K}\cite{FLR}.}.
However, in the Einstein-Hilbert truncation one must also take into account that $\Lambda(k)$ also \textit{runs}. If one considers the flow equations for $g(k)$ and the dimensionless cosmological constant $\lambda(k)$ [$\equiv k^{-2} \Lambda(k)$] one gets a non-trivial fixed point $(g*,\lambda^*)$ with $\lambda^* \in \Re^+$.
Therefore, as the energy scale grows, one gets $\Lambda(k)\rightarrow \infty$. Not surprisingly, when black holes are studied taking into account both the effects of the running Newton constant and the running cosmological constant one finds that the black hole singularity is reintroduced no matter the specific approach used to get the BH spacetime \cite{C&E,CKR,K&S,K&Sr}.

In this paper, we try to show that this conclusion can be avoided if one adopts the Unimodular Gravity (UG) approach. UG is a classical gravitational theory that imposes the metric of the spacetime to satisfy
\[
\sqrt{-det(g_{\mu\nu})}=\epsilon_0,
\]
where $\epsilon_0$ is a fixed scalar density. This has the effect of reducing the gauge symmetry from full spacetime diffeomorphism invariance (GR) to invariance only under diffeomorphisms that preserve this non-dynamical fixed volume element.
As a matter of fact, UG was initiated by Einstein in \cite{Ein} where he included the condition $det(g_{\mu\nu})=-1$ and he stressed that this condition was simply a \emph{choice of coordinates} made for convenience.
Indeed, if one starts with an unimodular action in which the unimodular condition is imposed from the beginning, then the resulting field equations correspond to the traceless Einstein equations \cite{Smolin}\cite{Fdez}. By using the Bianchi identity and the conservation of the energy-momentum, it can easily be shown that the traceless equations are equivalent to the full Einstein equations with a cosmological constant term, $\Lambda$, entering as an integration constant. Thus the equivalence of GR and UG at a classical level is made manifest. Nevertheless, the status of the cosmological constant is now different: while it was a coupling in the Einstein-Hilbert action, now it is just a constant of integration arising at the level of the equations of motion.
(Incidentally, it has been noted \cite{Smolin} that this is remarkable since it solves one of the cosmological constant puzzles, namely, the \textit{naturalness problem} which wonders why is $\Lambda$ not of the order of the \textit{natural} value $m_{Planck}^2$).

The new status of $\Lambda$ in Unimodular Gravity suggests that at a quantum level the differences between GR and UG could be drastic. In effect, recent studies \cite{Smolin}\cite{AGHM}\cite{Eich} indicate that the cosmological constant in Unimodular Gravity would be generated, but quantum corrections would not renormalize the classical value of the observable. In our view, this could also be a very interesting feature in favor of the UG approach since this would show that the theory could be devoid of the unwelcome infinite quantities arising from a running $\Lambda$.

In the case of UG the AS approach
leads to what is known as Unimodular Quantum Gravity (UQG) \cite{Eich}.
Our aim in this paper is to study black hole spacetimes by improving the classical unimodular solutions using the UQG approach.
We will analyze the characteristics of these spacetimes including their horizons, causal structure and, specially, their regularity.

The article is divided as follows. Section \ref{CUBH} is devoted to the study of the classical unimodular black holes. In sect.\ref{RNCUG} the running of the Newton constant as a function of the energy scale $G(k)$ is obtained in the framework of UQG. A proper cutoff identification provides us with its dependence on the \textit{radial distance} ``$\rho$''  to the black hole center $G(\rho)$. The quantum improved spacetime metric is deduced from this result in sect.\ref{IBHST} and its properties
are studied in sect.\ref{PQIS}. Finally, sect.\ref{conclu} is devoted to the conclusions.

\section{Classical Unimodular Black Hole}\label{CUBH}

\begin{figure}
\includegraphics[scale=1.3]{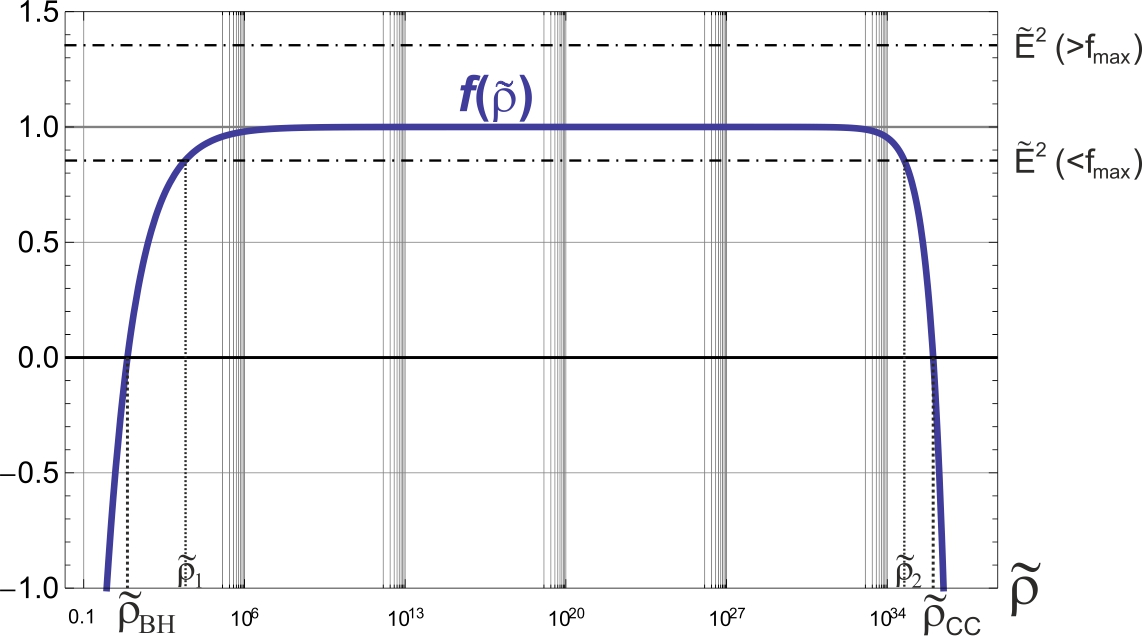}
\caption{\label{potential} A plot of $f$ as a function of $\tilde\rho\equiv 3 (G_0 m)^{-3}\, \rho$ for the particular case $\Lambda G_0^2 m^2 =3\cdot 10^{-24}$ (around the value corresponding to the biggest supermassive black hole currently known). A logarithmic scale has been used for the abscissa. In this case, the roots of $f$ provide us with the BH horizon $\tilde\rho_{BH}$ and the cosmological horizon $\tilde\rho_{CC}$. By using (\ref{frho}) one gets that the maximum of $f$ is $f_{max}=1-(9\Lambda G_0^2 m^2)^{1/3}$. This means that, in practice, even for this supermassive BH the function $f$ has its maximum very close to $1$ (specifically at $f_{max}=1- 3\cdot 10^{-8}$).  A particle describing a radial geodesic with an energy per unit mass $\tilde E$ satisfying $\tilde E^2>f_{max}$ will be able to travel from an initial $\rho_0$ satisfying $\rho_{BH}<\rho_0<\rho_{CC}$ either to $\rho=0$ or to $\rho\rightarrow\infty$. However, for $\tilde E^2<f_{max}$ a particle traveling in a radial geodesic will be bounded either by $0<\rho<\rho_1$ or by $\rho_2<\rho<\infty$, where $\rho_{1,2}$ are the solutions of $\tilde E^2=f$ satisfying $0<\rho_1<\rho_2$. (In all cases, when a particle is in the \textit{BH region} ($\rho<\rho_{BH}$) its radial position must decrease with time, while in the \textit{cosmological region} ($\rho>\rho_{CC}$) its radial position must increase with time).}
\end{figure}

In the framework of GR the solution for spherically symmetric black holes characterized by their mass $m$ and the presence of a cosmological constant $\Lambda$ is known as the Schwarzschild-de Sitter solution \cite{SdSs}.
A simple coordinate change allows us to write the Schwarzschild-de Sitter metric in unimodular form (see the appendix A) as
\begin{equation}\label{USdS}
ds^2 =-f(\rho) dt^2+\frac{1}{(3 \rho)^{4/3} f(\rho) } d\rho^2+ (3 \rho)^{2/3} \left[\frac{dx^2}{1-x^2}+(1-x^2) d\phi^2 \right],
\end{equation}
where
\begin{equation}\label{frho}
f(\rho)=1-\frac{2 G_0 m}{(3 \rho)^{1/3}}-\frac{\Lambda}{3} (3 \rho)^{2/3},
\end{equation}
$G_0$ is the usual Newton's gravitational constant and here (as in the rest of the paper) we are using units in which $c=\hbar=1$.
The coordinate $\rho$ has the dimensions of a volume and it is related to the area $A$ of the round spheres ($t$, $\rho$ constant) through the relationship $A^3=576\pi^3 \rho^2$.
This metric has trivial coordinate singularities at the solutions of $f(\rho)=0$ and at $x=\pm 1$. More importantly the metric has a well-known unavoidable curvature singularity at $\rho=0$.
The fact that $t$ does not appear in the metric coefficients (i.e., $t$ is a cyclic coordinate) implies that there is a killing vector $\vec k=\partial/\partial t$. A proper study reveals that this killing vector becomes light-like at $f=0$, what implies the existence of a horizon. %
It can be checked that there are two real positive roots of the cubic equation $f(\rho)=0$ whenever $0<9 \Lambda G_0^2 m^2<1$ which correspond to a black hole horizon $\rho_{BH}$ and a (classical) cosmological horizon $\rho_{CC}$.
These two roots can be written as
\begin{eqnarray}
\rho_{BH}&=&\frac{8}{3} \frac{\sin^3 \Psi}{\Lambda^{3/2}}\\
\rho_{CC}&=&\sqrt{\frac{3}{\Lambda^3}} \left(\cos\Psi -\frac{1}{\sqrt{3}}\sin \Psi  \right)^3,
\end{eqnarray}
where $\sin(3\psi)\equiv 3 G_0 m \sqrt{\Lambda}$.
It is easy to show \cite{PSR} that for $0<\rho<\rho_{BH}$ and for $\rho>\rho_{CC}$ the round spheres ($t$, $\rho$ constant) are closed trapped surfaces (i.e., $\rho_{BH}$ and $\rho_{CC}$ are also apparent 3-horizons).
A plot of $f$ showing the existence of these two horizons in a particular $0<9 \Lambda G_0^2 m^2<1$ case can be seen in fig.\ref{potential}.

Taking into account the current observational value for the cosmological constant $0\leq\Lambda \lesssim 10^{-52}\ m^{-2}$ and that the biggest observed (supermassive) black holes have Schwarzschild radius of the order of $R_S\lesssim 10^{14}\ m$ \cite{Ghise}, one has $0<9 \Lambda G_0^2 m^2< 10^{-24}\ll 1$, so that the case with two horizons seems to be the only relevant one in practical terms. Moreover, in this case it is easy to check that the value obtained for the position of the BH horizon corresponds, within a very good approximation, to that of the Schwarzschild black hole horizon ($\Lambda=0$) since
\begin{equation}\label{robh}
\rho_{BH} \simeq \frac{8}{3} G_0^3 m^3 (1+4 \Lambda G_0^2 m^2)  \simeq \frac{8}{3} G_0^3 m^3.
\end{equation}
On the other hand, the position of the cosmological horizon would be approximately the position that would be obtained for a de Sitter cosmological horizon ($m=0$) since
\begin{equation}\label{roc}
\rho_{CC} \simeq \frac{1}{3} \left(\sqrt{\frac{3}{\Lambda}}-G_0 m \right)^3\simeq \sqrt{\frac{3}{\Lambda^3}}.
\end{equation}

\section{Running Newton constant in UQG}\label{RNCUG}

As stated in the introduction, UQG aims to find an UV complete theory of Unimodular gravity in the context of Asymptotic Safety. One way to achieve this is
by imposing the condition $\sqrt{-det(g_{\mu\nu})}=\epsilon_0$
in the action and the path integral, what reduces the dynamical variables (when compare to the GR case) \cite{Unruh}\cite{Eich}. A summary of the procedure would be as follows.
The effective average action $\Gamma_k$ for gravity satisfies the following Wetterich-type \cite{Wett} Functional Renormalization Group Equation (FRGE) \cite{Reuter}
\begin{equation}\label{tgammak}
\partial_t\Gamma_k=\frac{1}{2} \mbox{STr} \left[(\Gamma^{(2)}_k+R_k)^{-1} \partial_t R_k\right],
\end{equation}
where $\partial_t=k \partial_k$, STr is the super-trace (over all fields, indices and an integral over spacetime), $\Gamma^{(2)}_k$ is the second functional derivative of $\Gamma_k$ with respect to the dynamical fields and $R_k$ is an infrared mass-like regulator  that suppresses IR modes with $p^2<k^2$ in the generating functional.
In order to follow the background field formalism \cite{Abbott} one has to split the metric into background and fluctuation field and adapt this to the unimodular setting.
In addition, one has to introduce a background field gauge fixing condition $S_{gf}$ and a Faddeev-Popov ghost sector $S_{gh}$ in the effective average action with both $S_{gf}$ and $S_{gh}$ adapted to the unimodular condition \cite{Eich}.

If one assumes that there is a set of basis functionals spanning the theory space, one could write $\Gamma_k$ as a linear combination of an infinite number of the basis functionals, being the coefficients the scale dependent couplings. Then, by using the FRGE one would obtain a system of infinitely many coupled differential equations that would be too hard to solve in general. The usual way out is to restrict the analysis to a finite-dimensional subspace of the full theory space.
Here we are considering a truncation with the form
\begin{equation}\label{gammak}
\Gamma_k=-\frac{k^2}{16 \pi g(k)} \int d^4 x\, \epsilon_0\, R+ S_{gf}+S_{gh}.
\end{equation}
By using (\ref{gammak}) together with a proper regulator $R_k$ \cite{LitimR} in (\ref{tgammak}) one gets
the flow equation for the dimensionless Newton constant $g$ in UQG \cite{Eich}
\begin{equation}\label{flow}
\partial_t g=\beta(g),
\end{equation}
where
\[
\beta(g)=2 g+\frac{g^2 \ A_1}{A_2-A_3 g}
\]
and $A_i$ ($i=1,2,3$) are constants\footnote{See sect. III of \cite{Eich}: $A_1=3 (1300 -309 \sqrt{13}-325 \sqrt{17})\ (<0)$, $A_2=936\pi$ and $A_3=1625$.}.
This can also be rewritten for our purposes as
\begin{equation}\label{beta}
\beta(g)=2 g \left(\frac{1-\omega g}{1-B g} \right),
\end{equation}
where $\omega\equiv (2 A_3-A_1)/(2 A_2)$ and $B\equiv A_3/A_2$.
The flow equation has two critical points. In one hand, the Gaussian fixed point $g^*=0$ and, on the other hand, the more interesting non-Gaussian fixed point $g^*=1/\omega$ (which suggests that UQG could be an UV complete quantum theory of gravity).

The integration of (\ref{flow}) using (\ref{beta}) between a reference energy scale $k_0$ and the energy scale $k$ leads us to
\begin{equation}\label{gk}
\frac{g(k)}{k^2 (1-g(k) \omega)^{\zeta}}=\frac{g(k_0)}{k_0^2 (1-g(k_0) \omega)^{\zeta}},
\end{equation}
where we have defined $\zeta\equiv 1-B/\omega$.
An approximate analytical expression for $g(k)$ can be obtained if we take into account that, according to the numerical values in \cite{Eich},
\begin{equation}\label{omom}
\zeta^{-1}=1-2\frac{A_3}{A_1}\simeq 2.
\end{equation}
(It is remarkable that previous works
in the framework of QEG and, specifically, in the Einstein-Hilbert truncation \cite{Reuter}\cite{B&R}\cite{Eich} had found $\zeta\simeq 1$).
If one now uses the approximation (\ref{omom}) in order to get an analytical expression for the running $G$ from (\ref{gk}), one gets
\[
G(k)=G_0 \left(\frac{G_0 k^2 \omega-\sqrt{G_0^2 k^4 \omega^2-4 G_0 k_0^2+4}}{2 (G_0 k_0^2 \omega-1)} \right),
\]
where the definition of $g(k)$ [$\equiv G(k) k^2$] and $G(k_0)= G_0$ have been used.
Of course, the energy scales in which the Newton constant has been experimentally determined  in the laboratory are very small with respect to the Planckian energy scales so that we can take $k_0\approx 0$ and identify $G(k_0=0)\equiv G_0$ from which we get
\begin{equation}\label{Gk}
G(k)=G_0 \left(\frac{\sqrt{G_0^2 k^4 \omega^2+4}-G_0 k^2 \omega}{2} \right).
\end{equation}

The infrared limit of this running gravitational coupling provide us with the expected result
\[
G(k\rightarrow 0)\rightarrow G_0,
\]
while the ultraviolet limit  takes the (independent of $G_0$) form
\[
G(k)\simeq \frac{1}{k^2 \omega}+\mathcal{O}(k^{-3}),
\]
so that the gravitational coupling weakens with increasing energy scales and, eventually, $G(k\rightarrow\infty)=0$. This is precisely the behaviour conjectured in \cite{Poly} and already obtained in the framework of QEG \cite{B&R}.

\subsection{Cutoff identification}
Let us now apply the results for the running $G$ in UQG to the case of spherically symmetric black holes. We will quantum improve the spacetime describing the BH by assuming that the classical coupling $G_0$ is replaced by a running coupling $G$, which could depend on the coordinates and the parameters ($m$ and $\Lambda$) characterizing the spacetime (see, for example, \cite{B&R}\cite{B&K}\cite{FLR}\cite{K&S}\cite{K&Sr}\cite{KCRS}). The search for a static spherically symmetric quantum improved spacetime imposes that $G$ should be independent of $t$ and the angular coordinates $\theta$ and $\phi$. In this way, we are essentially searching for a running coupling $G=G(\rho; m,\Lambda)$. Now, since the renormalization group has provided us with a momentum dependent $G(k)$, we still require an identification between the momentum scale $k$ and the coordinate $\rho$, what can be formally written as
\begin{equation}\label{kxid}
k(\rho; m, \Lambda)=\frac{\xi}{d(\rho; m, \Lambda)},
\end{equation}
where the (dimensionless) numerical value $\xi$ should be fixed later and $d(\rho; m, \Lambda)$ is the \textit{distance scale} which provides the relevant cutoff for the Newton constant when a test particle is located at a given $\rho$. Clearly, we need $d$ to have the dimensions of a ``distance''. It also should be an invariant, i.e., independent of the chosen coordinates. In this way, a common choice in the literature has been to define \cite{B&R}\cite{K&Sr}\footnote{As far as I am aware, in the literature one can only find one alternative definition useful for BH spacetimes \cite{F&L}. Nevertheless, it provides similar qualitative results.}
\begin{equation}\label{dro}
d=\int_{\mathcal C} \sqrt{|ds^2|},
\end{equation}
where, there is still an ambiguity in identifying the correct spacetime curve $\mathcal C$ in which the integration is carried out from $\rho=0$ to the specific value of $\rho$. Let us remark the use of a modulus in the square root in (\ref{dro}) since this \textit{distance} does not even impose a causal character to the curve $\mathcal C$ (which could even vary piecewise).

\begin{figure}
\includegraphics[scale=.8]{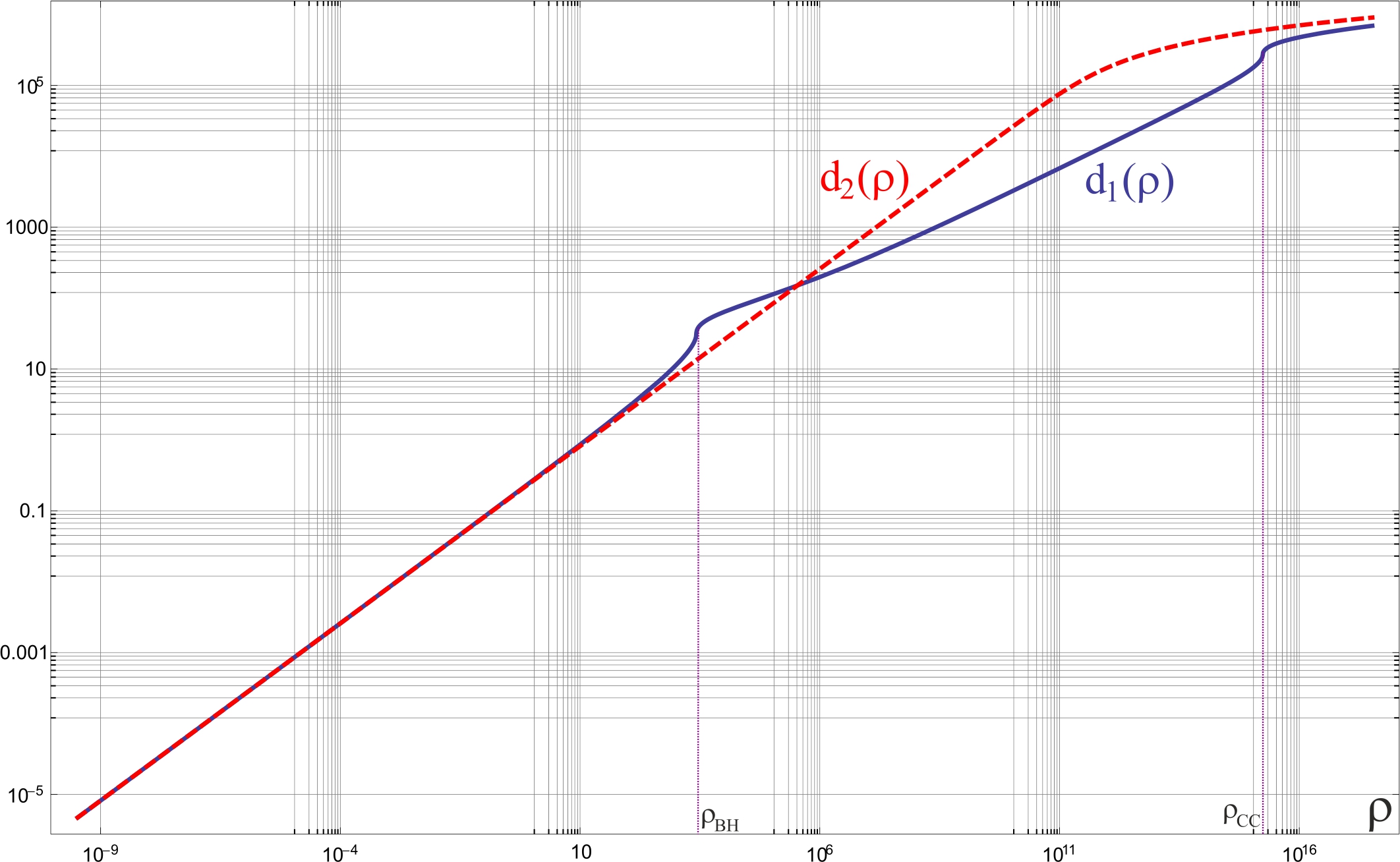}
\caption{\label{figd} A logarithmic plot of the distance scales $d_1(\rho)$ and $d_2(\rho)$ in natural units. Note that $d_1(\rho)$ is a continuous function at the horizons $\rho_{BH}$ and $\rho_{CC}$. Note also how the \textit{short distance} and the \textit{long distance} behaviour of $d_1(\rho)$ and $d_2(\rho)$ coincide. Here we have chosen $m=10\ m_{Planck}$, $\Lambda G_0^2 m^2=10^{-8}$, but the qualitative features are independent of the parameters as long as the two horizons exist.}
\end{figure}

In order to compare our results with the results in the QEG approach we will now consider the usual curves analyzed in that approach when dealing with static spherically symmetric BHs.
Consider, first, the so-called \textit{straight radial curve}\cite{B&R}\cite{K&Sr}   $\mathcal C_1$ parametrized with $\kappa$:  $x^\mu(\kappa)=\{t(\kappa)=t_0,\rho(\kappa)=\kappa,x(\kappa)=x_0,\phi(\kappa)=\phi_0\}$.
By using (\ref{USdS}) we have (from now on we will not make explicit the dependence on the parameters $m$ and $\Lambda$)
\begin{equation}\label{dsc1}
d_1(\rho)=\int_0^\rho \frac{d\rho}{(3 \rho)^{2/3}\sqrt{|f(\rho)|}}.
\end{equation}
The integrand in this expression diverges at the horizons (were the causal character of $\mathcal C_1$ changes). However, it can be shown that the integral provide us with a continuous function $d_1(\rho)$ (see fig. \ref{figd}).
The approximate analytic behaviour of the distance scale (\ref{dsc1}) for $\rho\simeq 0$ is simply
\begin{equation}\label{shortd1}
d_1(\rho)\simeq \sqrt{\frac{2\rho}{3 G_0 m}}.
\end{equation}
On the other hand, for large $\rho$ ($\rho\ggg\rho_{CC}$) the asymptotic analysis of (\ref{dsc1}) provide us with\footnote{Alternatively, one can get this result by considering that the parameter $\Lambda$ becomes relevant at these distances, while $m$ becomes irrelevant (see the appendix B).}
\begin{equation}\label{longd1}
d_1(\rho)\simeq \frac{1}{\sqrt{3 \Lambda}} \ln(\Lambda^{3/2} \rho).
\end{equation}

The other usual curve considered in the QEG literature \cite{B&R}\cite{K&Sr} is the radial geodesic $\mathcal C_2$ described by a test particle of mass $\mu$ with a proper time per unit mass $\bar\tau$: $t=t(\bar\tau)$ and $\rho=\rho(\bar\tau)$ (see, for instance, \cite{gravit} for the general procedure).
For this trajectory, and taking into account that the coordinate $t$ is cyclic,
one has $p_t=-E$, $p_\rho=g_{\rho\rho} d\rho/d\bar\tau$, $p_x=0$ and $p_\phi=0$. Then, $g^{\alpha\beta}p_\alpha p_\beta=-\mu^2$ provides us (after considering the different possibilities for the trajectories) with the formal compact expression
\begin{equation}\label{dist}
\Delta \tau=\int_0^\rho \frac{d\rho}{(3\rho)^{2/3}\sqrt{\tilde E^2-f(\rho)}},
\end{equation}
where $\tilde E\equiv E/\mu$ is the energy per unit mass and $\tau\equiv \bar\tau \mu$ is the particle proper time. [Note that the signs and the order of the integration limits in the expression have been chosen to provide the correct results in any situation: First, when we want to compute the time taken for a particle to travel from $\rho<\rho_{CC}$ to $\rho=0$ (necessarily in this order for the particle) or, second, when we want to compute the time to ``travel between $\rho>\rho_{CC}$ and $\rho=0$''. In this last case, we must take the time taken from a particle to travel from an intermediate $\rho^*<\rho_{CC}$ to $\rho=0$ and add the time taken for a particle to travel from $\rho_*$ to $\rho$ ($> \rho_{CC}$). This is already taken into account in (\ref{dist})]. Considering now that for the radial geodesic $ds^2=-d\tau^2$, we directly get the distance scale in this case from its definition (\ref{dro}). A specially simple analytical expression can be found for the particular case $\tilde E^2=1$ taking the form
\begin{equation}\label{distscale}
d_2(\rho) =\frac{2}{\sqrt{3 \Lambda}}\, \mbox{arsinh} \sqrt{\frac{\Lambda \rho}{2 G_0 m}}.
\end{equation}
For $\rho\simeq 0$ this can be approximated by
\begin{equation}\label{shortd2}
d_2(\rho)\simeq \sqrt{\frac{2\rho}{3 G_0 m}},
\end{equation}
while for large distances ($\rho\ggg\rho_{CC}$)
\begin{equation}\label{longd2}
d_2(\rho)\simeq \frac{1}{\sqrt{3 \Lambda}} \ln(\Lambda^{3/2} \rho).
\end{equation}
Is is easy to check that even for $\tilde E^2\neq 1$ the
expression (\ref{shortd2}) is a good approximation of (\ref{dist}) for small $\rho$'s. Likewise, for large distances and $\tilde E^2\neq 1$ the
expression (\ref{longd2}) is a good approximation of (\ref{dist}) \footnote{Of course, we are assuming that the energy per unit rest mass of the particle $\tilde E$ is chosen big enough to let the particle reach such distances, i.e., to let it classically traverse the potential barrier (see figure \ref{potential}), what requires $\tilde E^2>f_{max}$. This will always be true, for example, for $\tilde E^2>1$.}.

Comparing the results obtained by using the curves of type $\mathcal C_1$ and $\mathcal C_2$ we see that the qualitative behaviour of their distance scales coincide. Not only does the long distance behaviour ((\ref{longd1}) and (\ref{longd2})), but also does the short distance behaviour ((\ref{shortd1}) and (\ref{shortd2})), where the quantum modifications are expected to be relevant. (See also figure (\ref{figd})).
Therefore, since we have the analytical expression (\ref{distscale}), it would seem sensible to consider, from now on, (\ref{distscale}) as our \textit{interpolating distance scale} \cite{B&R}\cite{K&Sr}. In this way, one would expect that the results obtained by using it to be \emph{qualitatively} correct.

\subsection{The running $G(\rho)$}

Let us now write $G(\rho)$ by using (\ref{Gk}) and (\ref{kxid}) as
\begin{equation}\label{Grho}
G(\rho)=G_0\left(\sqrt{1+\frac{G_0^2 \tilde\omega^2}{4 d^4(\rho)}}-\frac{G_0\tilde\omega}{2 d^2(\rho)} \right),
\end{equation}
where the new dimensionless constant $\tilde\omega\equiv \omega\xi^2$ and $d(\rho)$ is given by (\ref{distscale}).
A relevant fact with regard to $\tilde\omega$ is that it carries the quantum
modifications to the gravitational coupling. In effect, $\tilde\omega=0$ would turn off the running of $G$. Moreover, if we explicit Planck's constant we see that  $\tilde\omega\propto \hbar$.
In principle, the precise value of $\tilde\omega$
can be found by comparison with the standard perturbative quantization of
Unimodular Gravity. Previous results in the perturbative quantization of General Relativity (see \cite{BBDH} and references therein) show that $\tilde \omega =167 \hbar/(30\pi) \sim \hbar$ \cite{B&R}\cite{TFL} so that, since the qualitative properties of $G$ do not rely on its precise value (as long as it is strictly positive), we will assume for numerical computations that $\tilde \omega \simeq \hbar$.

For small distances
the expression (\ref{Grho}) reduces to
\[
G(\rho) \simeq \frac{2 \rho}{3\tilde\omega G_0 m},
\]
so that, on the one hand,
\[G(\rho\rightarrow 0)=0
\]
(which points toward a weakening of gravitational effects in the innermost region of the BH) and, on the other hand, we see that the role played by $\Lambda$ at small distances is negligible.

\begin{figure}
\includegraphics[scale=.7]{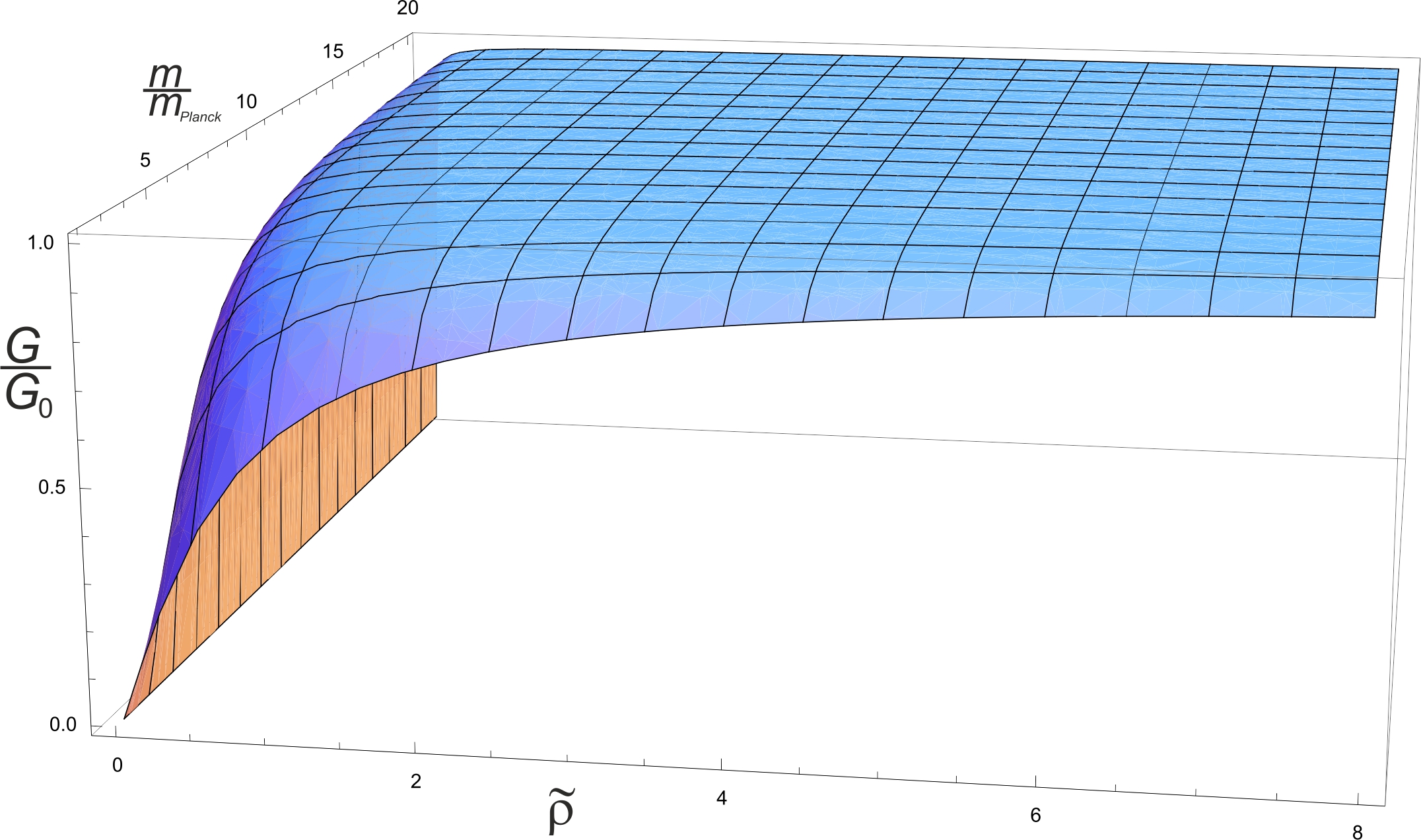}
\caption{\label{runningG} A plot of $G/G_0$ as a function of $\tilde\rho\equiv 3 (G_0 m)^{-3}\, \rho$ and the BH mass ($/m_{Planck}$). As can be seen, for a given BH mass,  $G$ monotonously increases from its minimum value $G(\rho=0)=0$ towards its maximum (classical) value $G_0$.
}
\end{figure}

For large distances ($\rho\ggg \Lambda^{-3/2}$) we get
\begin{equation}\label{Gbig}
G(\rho) \simeq G_0- \frac{3 \tilde\omega G_0^2 \Lambda}{2 [\ln(\Lambda^{3/2} \rho )]^2 }=
G_0 \left(1-\frac{3 \tilde\omega \Lambda l_{Planck}^2 }{2 [\ln(\Lambda^{3/2} \rho)]^2} \right),
\end{equation}
where $l_{Planck}$ is Planck's length. Therefore, we find the expected result
\[
G(\rho\rightarrow \infty)=G_0
\]
and we see that $\Lambda$ does play a role (by weakening $G$) in the leading term for the quantum corrections of the gravitational constant at large distances.
A graph of the running G obtained can be seen in figure \ref{runningG}.
%

\section{Improved black hole spacetime}\label{IBHST}

In classical Unimodular Gravity (as in classical General Relativity) the metric (\ref{USdS}) has a meaning even in the absence of test particles to probe it. As we have seen, in the Asymptotic Safety approach, the presence of the test particle defines a physically relevant distance scale $d(\rho)$ which enters into the cutoff for the running of $G$. If one assumes that the leading quantum effects in the system consist of a position dependent renormalization of Newton's gravitational constant appearing in the unimodular classical metric (see, for example, \cite{B&R}\cite{B&K}\cite{FLR}\cite{K&S}\cite{K&Sr}\cite{KCRS}), we will have a quantum improved line element for the spacetime of the form (\ref{USdS})
\begin{equation}\label{iUSdS}
ds^2 =-f_I(\rho) dt^2+\frac{1}{(3 \rho)^{4/3} f_I(\rho) } d\rho^2+ (3 \rho)^{2/3} \left[\frac{dx^2}{1-x^2}+(1-x^2) d\phi^2 \right],
\end{equation}
with
\begin{equation}\label{ifrho}
f_I(\rho)=1-\frac{2 G(\rho) m}{(3 \rho)^{1/3}}-\frac{\Lambda}{3} (3 \rho)^{2/3},
\end{equation}
and $G(\rho)$ given by (\ref{Grho}).

As a consequence, it is easy to check that the classically diverging $f(\rho\rightarrow 0)$ behaviour is now replaced around $\rho=0$ by\footnote{Clearly, this development and what follows is correct if $\tilde\omega\neq 0$. I.e., we are describing now a fully quantum effect. Of course, there is not \textit{de Sitter core} in the classical case, but a divergence of $f$.}
\begin{equation}\label{fpeq}
f_I(\rho)\simeq 1-\left(\frac{4}{9 G_0 \tilde\omega}+\frac{\Lambda}{3} \right) (3 \rho)^{2/3},
\end{equation}
which corresponds to a \textit{de Sitter core} [by considering that in the exact unimodular de Sitter solution one has $f_{dS}(\rho)=1-(\Lambda_{eff}/3) (3 \rho)^{2/3} $], with an effective \textit{cosmological} constant
\begin{equation}\label{lameff}
\Lambda_{eff}=\frac{4}{3 G_0 \tilde\omega}+\Lambda .
\end{equation}
Since the first term in the rhs is of the order of $l_{Planck}^{-2}$, clearly, the role played by $\Lambda$ in the interior of the BH will be negligible. Thus, the behaviour is very close to that found in \cite{B&R}, where the Quantum Einstein Gravity approach without cosmological constant provides (\ref{lameff}) (with $\Lambda=0$).

\section{Properties of the quantum improved solution}\label{PQIS}

\subsection{Regularity of the solution}

Our study of $f_I(\rho\simeq 0)$ strongly suggests that, contrary to the classical case, the improved spacetime will not have a scalar curvature singularity at $\rho=0$. In principle, to strictly prove this (and given that we are dealing with the \textit{center} of a spherically symmetric spacetime) it would be enough to check two algebraically independent curvature invariants \cite{F&T}. The computations for this specific case show that, in fact, here there is only one algebraically independent non-zero curvature invariant for this spacetime at $\rho=0$ which can be taken to be the Ricci scalar $\mathcal R$. Around $\rho=0$ the Ricci scalar takes the form
\[
\mathcal R=4 \left(\frac{4}{3 G_0 \tilde\omega}+\Lambda \right)+ \mathcal O (\rho),
\]
so that it is finite and, therefore, there will not be scalar curvature singularities in this quantum improved spacetime.

\subsection{Horizons}

There is a killing vector $\vec k=\partial/\partial t$ in the improved spacetime since the coordinate $t$ does not appear in the metric coefficients of (\ref{iUSdS}).
This vector becomes lightlike if there is a $\rho_h$ such that $f_I(\rho_h)=0$. Then $\rho=\rho_h$ would define a spherically symmetric lightlike hypersurface called a killing horizon that can be shown \cite{PSR} to be also an apparent 3-horizon.
While (at a theoretical level) in the classical Schwarzschild-de Sitter solution one can choose the parameters such that the solution would not have any horizon ($9\Lambda G_0^2 m^2 >1$), in the improved solution this situation will not be possible and there will always be one or more horizons. This is a consequence of the fact that
$f_I$ is a continuous function in $\Re^+$ with $f_I(\rho\rightarrow 0)=1>0$ and $f_I(\rho\rightarrow \infty)=-\infty<0$.

On the other hand, it is not possible to obtain an exact analytical solution for the zeros of $f_I(\rho)=0$ in the general case. However, it is possible to study the zeros numerically and to find some approximate values (by using the observational limits of the parameters commented in sect.\ref{CUBH}). For example, (\ref{Gbig}) tell us that the large distance quantum corrections to the classical Schwarzschild-de Sitter black hole solution should be negligible. Therefore, according to the observational limits for $\Lambda$ and $m$ there should be an \textit{improved} cosmological horizon $\rho_C$ solution of $f_I=0$ around the classical value $\rho_{CC}$ (\ref{roc}). By finding an approximate expression for $f_I$ around this value and searching for its zero we find the improved cosmological horizon to be approximately given by
\[
\rho_C \simeq \rho_{CC} (1+\alpha \tilde\omega G_0^2 m \Lambda^{3/2})
\]
where $\alpha$ is a positive constant\footnote{The expansion of the solution in terms of the small dimensionless parameters $X\equiv \Lambda G_0^2 m^2$ and $Y\equiv \tilde\omega/(G_0 m^2)$ provide us with $\alpha=6 \sqrt{3}/(\ln 3)^2 \simeq 8.61$.}. In this way, we get that the quantum correction slightly enlarges the classical cosmological horizon $\rho_{CC}$.

\begin{figure}
\includegraphics[scale=1.2]{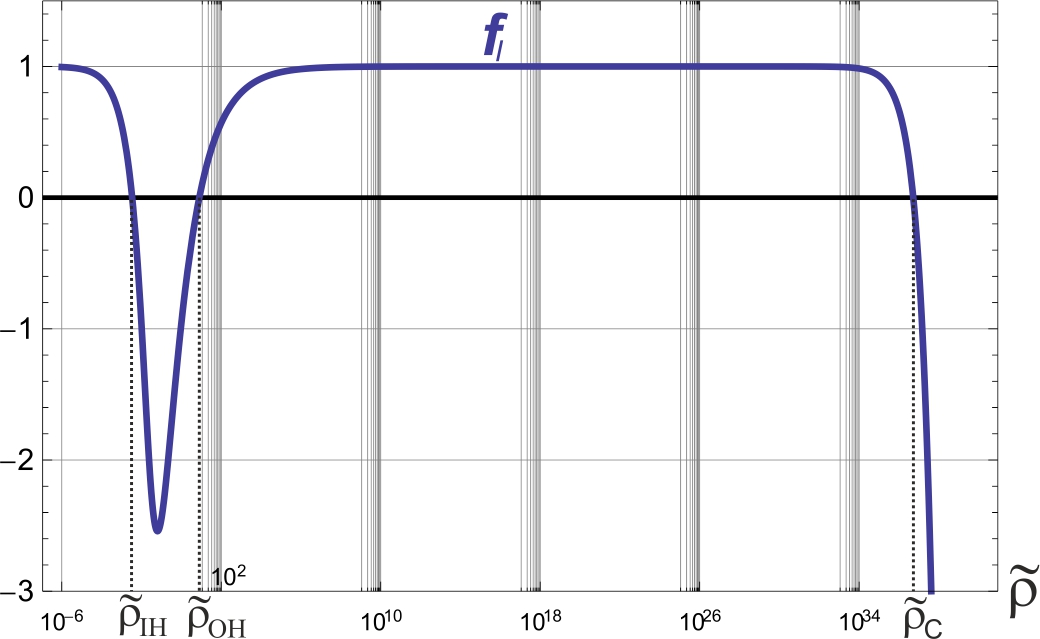}
\caption{\label{fiM} A plot of $f_I$ as a function of $\tilde\rho\equiv 3 \rho/(G_0 m)^{3}$ for a particular case with $\Lambda G_0^2 m^2 =3\cdot 10^{-24}$ and $m\gg m_{Planck}$ (i.e., similar values as those used for the plot of the classical case in fig.\ref{potential}). A logarithmic scale has been used for the abscissa. In this quantum improved case three horizons (corresponding to the zeros of $f_I$) appear: The cosmological horizon $\tilde\rho_C$, the outer horizon $\tilde\rho_{OH}$
and the \textit{new} inner horizon $\tilde\rho_{IH}$.
}
\end{figure}

On the other hand, for astrophysical size black holes ($m\ggg m_{Planck}=G_0^{-1/2}$)
one expects a very small quantum correction in the position of the BH horizon (now $\rho_{OH}$) with respect to its classical value $\rho_{BH}$ (\ref{robh}). In effect, we can approximately solve $f_I=0$ around the classical value to get
\[
\rho_{OH}\simeq \rho_{BH} \left(1-\frac{27}{32}  \frac{\tilde\omega}{G_0 m^2} \right),
\]
where we see that now the quantum correction slightly shrinks the classical horizon.
The subindex (OH) here stands for \textit{outer horizon}. The nomenclature is due to the fact that, for these astrophysical black holes, one expects the quantum effects to also create another \textit{inner horizon} (IH). The reason is that in the $m\ggg m_{Planck}$ case and for $\rho$'s slightly smaller than $\rho_{OH}$ one will have a black hole region with $f_I<0$ ($df_I/d\rho\rfloor_{\rho_{OH}}>0$),
however $f_I(\rho=0)=1>0$ and, since $f_I$ is a continuous function in $\Re^+$, it should have at least another inner zero.
The position of the inner horizon $\rho_{IH}$ around $\rho=0$ can be very well approximated for the $m\ggg m_{Planck}$ case by using the expression for $f_I$ in (\ref{fpeq}) and demanding it to be zero. The result is
\[
\rho_{IH}\simeq\frac{9}{8} (G_0 \tilde\omega)^{3/2} \left(1-\frac{9}{8} G_0 \tilde\omega \Lambda \right).
\]
Note that, as expected, the role of the cosmological constant
in determining the position of the inner horizon is negligible. More importantly, for these large masses and within this approximation, the position of the inner horizon turns out to be independent of the precise value of the black hole mass.
In fig.\ref{fiM} we plot the quantum improved function $f_I$ for this $m\gg m_{Planck}$ case in order to show the position of the three horizons when the parameters take similar values to those used in fig.\ref{potential} (for the classical case).

\begin{figure}
\includegraphics[scale=1.2]{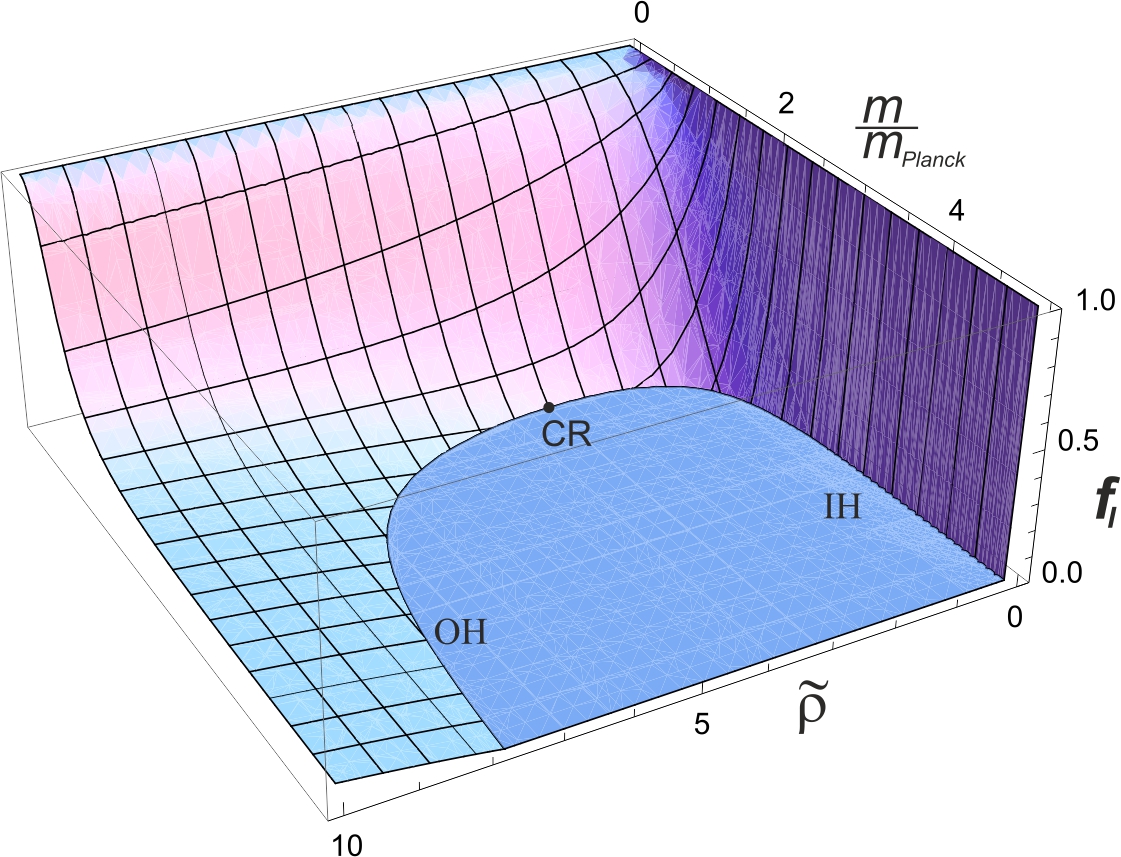}
\caption{\label{f3D} A plot of $f_I$ centered in the innermost BH region for different BHs with masses of the order of the planckian mass. Only the positive values of $f_I$ are shown so that it is easy to identify the (flat) region where the 2-spheres are closed trapped surfaces ($f_I <0$) and the locus of the horizons ($f_I=0$, i.e., the boundary of the flat region). In particular, we can see how the outer (OH) and inner (IH) horizons converge into a single \textit{critical} horizon (CR) in $\tilde\rho=\tilde\rho_{cr}$ when the considered BH mass has the value corresponding to the critical mass $m_{cr}$.}
\end{figure}

The scenario is different for planckian size black holes since
a numerical computation shows that
there exists a \textit{critical mass} $m_{cr}$ for which there is only a single \emph{black hole} horizon (see figure \ref{f3D}) (coexisting with the cosmological one).
In other words, as one considers smaller BH masses, two of the zeros of $f_I$ are transformed into a double zero when the BH reaches the critical mass.
This critical mass can be numerically computed when one chooses particular values for the parameters. For example, if $\Lambda=10^{-52} m^{-2}$ and $\tilde\omega\simeq\hbar$ we get
$m_{cr} \simeq 1.5\, m_{Planck}$ with the black hole horizon situated at $\rho_{cr} \simeq 3.18\, l_{Planck}^3$.
Finally, when the mass is smaller than the critical mass there is no \emph{black hole} horizon, but only the cosmological horizon (fig.\ref{f3D}).

\subsection{Causal structure}

\begin{figure}
\includegraphics[scale=1]{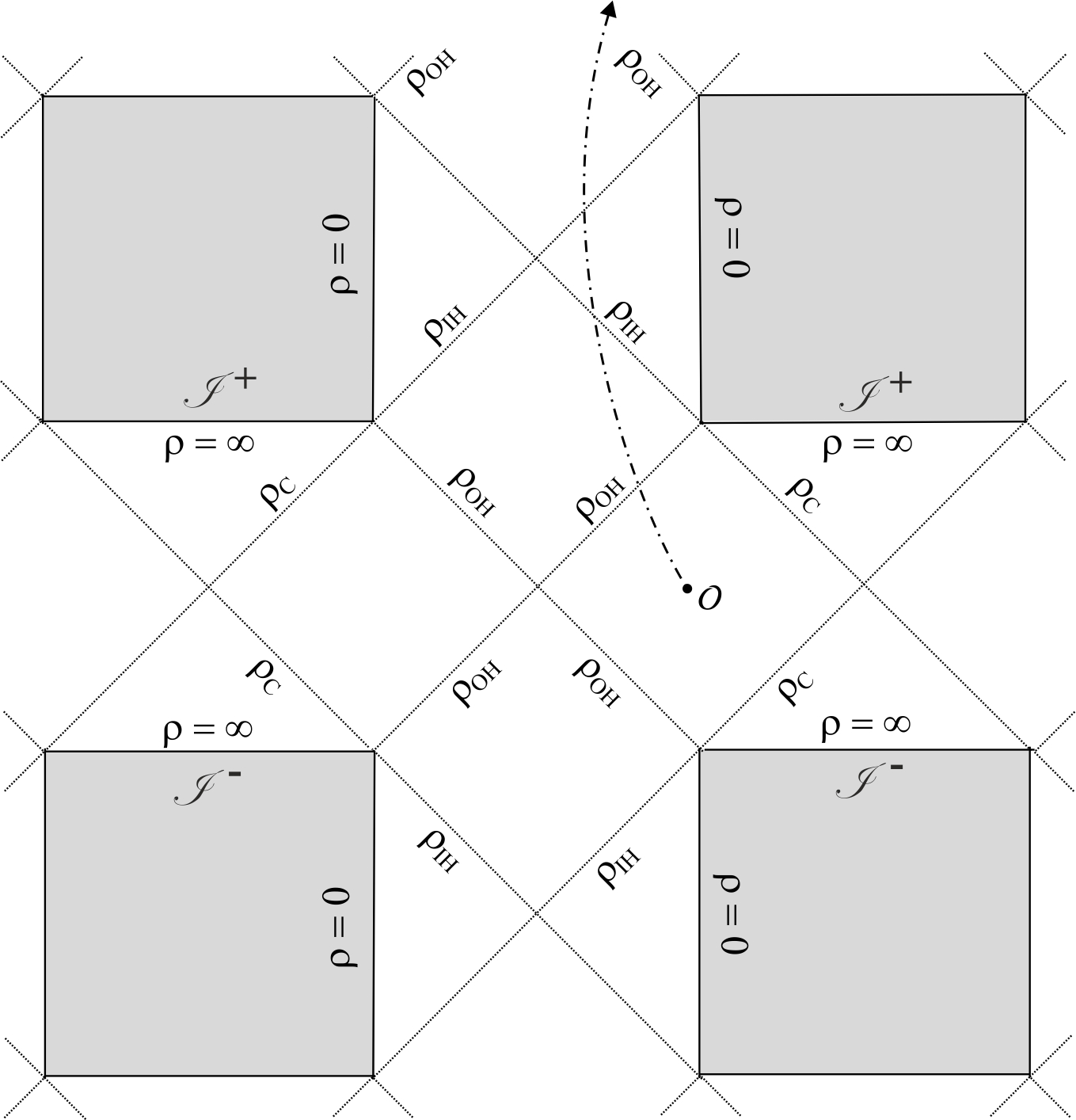}
\caption{\label{IBHPenrose} A portion of the conformal diagram for the improved black hole spacetime with $m>m_{cr}$ and assuming $0<\Lambda\lesssim 10^{-52} m^{-2}$. The grey regions are not part of the spacetime. A test particle $\mathcal O$ initially between the outer horizon of the black hole $\rho_{OH}$ and the cosmological horizon $\rho_{C}$ can traverse $\rho_{OH}$. In the region between this horizon and the inner horizon $\rho_{IH}$ the round spheres are closed trapped surfaces so that the particle has to reach the inner horizon. Were this horizon stable, the particle could not reach any singularity. In particular, the particle could reach the center of symmetry $\rho=0$, where the curvature is finite, and continue its travel without any disruption. In principle, it could traverse the interior through the (white hole) outer horizon $\rho_{OH}$ in the upper part of the drawing.}
\end{figure}

A portion of the global causal behaviour of the improved black hole spacetime with $m>m_{cr}$ is depicted in the conformal diagram of fig.\ref{IBHPenrose}. The complete maximally extended conformal diagram of the spacetime consists of an infinite repetition of this pattern.

This improved black hole shares with the classical Reissner-Nordtr\"{o}m spacetime with $Q^2<G_0 M^2$ the feature of having an inner and an outer (lightlike) BH horizon. However, they disagree in the fact that for the improved black hole there is a non-singular center of symmetry while in the RN case the would-be center of symmetry is not even part of the spacetime (since there is a curvature singularity).

According to (\ref{Gbig}) and (\ref{ifrho}), for large distances the quantum gravitational effects become negligible and the cosmological constant dictates the behaviour of the spacetime metric. In this way, the causal asymptotic behaviour of the improved black hole becomes similar to that in the classical de Sitter case. In particular, the causal structure has spacelike conformal infinities $\mathscr{I}^\pm$ at $\rho=\infty$.

\section{Conclusions}\label{conclu}

In this paper we have tried to show that Asymptotic Safety could be able to provide us with quantum improved \emph{non-singular} black hole spacetimes. Specifically, we have seen that this could be achieved by following an UQG framework in which the cosmological constant $\Lambda$ is generated, but does not \textit{run}  \cite{Smolin}\cite{AGHM}\cite{Eich}.
Furthermore, we have shown that then there is a running gravitational coupling $G$ that tends to zero as the energy scale increases so that the gravitational coupling weakens at high energies.

In order to obtain the spherically symmetric black hole spacetimes incorporating the UQG improvements, we have first found a proper \textit{interpolating distance scale} that has allowed us to deduce the qualitative behaviour of the gravitational coupling $G$ with respect to a chosen areolar coordinate $\rho$. The effective quantum spacetime describing black holes characterized by their mass and the presence of a cosmological constant (\ref{iUSdS}) has then been found by ussing the common procedure (\cite{B&R}\cite{B&K}\cite{FLR}\cite{K&S}\cite{K&Sr}\cite{KCRS}) of improving the corresponding classical unimodular metric ($G_0\rightarrow G(\rho)$).
We have shown that, indeed, this improved spacetime is devoid of singularities. This contrasts with similar computations in the framework of QEG where the existence of a non-gaussian fixed point for the (necessarily) running $\Lambda$ leads to singular BH spacetimes \cite{C&E,CKR,K&S,K&Sr}. Nevertheless, it has to be taken into account that the method of quantum improving a classical spacetime is expected to provide us only with good \emph{qualitative} results far from the Planckian regime. In this way, it can be said that the method only \emph{suggests} the avoidance of the singularity, what is better grounded in a vanishing running $G$ and, especially, in the existence of the constant $\Lambda$ suggested by UQG.

If we assume $\Lambda>0$ (and within its current observational limit) and if the BH mass is bigger than a critical mass $m_{cr}$ (of the order of Planck's mass) then the improved spacetime possesses three horizons: A cosmological horizon, an outer black hole horizon and an inner black hole horizon. The cosmological and the outer horizons can be considered as quantum improved versions of the corresponding horizons in the classical case. In fact, we have seen that if the BH mass satisfies $m\gg m_{cr}$ then the quantum correction to their positions is negligible. In contrast, the inner horizon is a truly new feature of the quantum improved spacetime.

The horizon structure changes if the BH mass equals the critical mass. In this case there is a cosmological horizon and a single BH horizon. Finally, if the BH mass is smaller than the critical mass only the cosmological horizon exists.

The interesting causal structure for the case of BHs with masses bigger than the critical mass has been shown in fig.\ref{IBHPenrose}.
It cannot be forgotten that this structure will be modified by other unavoidable physical effects. In particular, the existence of the horizons will be related to the emission of Hawking radiation
with the subsequent modification of the BH mass. Therefore, the horizons will not be able to remain static (and lightlike).
On the other hand, for similar reasons to those found in \cite{TorresIns} one expects that the inner horizon could be unstable under the effect of perturbations on it (such as radiation and particles entering the BH), so that a study of its behaviour would be required.

Finally, note that the results presented here come from the existence of a fixed point through the use of the truncation (\ref{gammak}). However, some studies for QEG \cite{C&E}\cite{Falls} and for UQG \cite{Eichf} confirm that one can consider other relevant terms and still show the existence of a non-trivial fixed point of the theory's renormalization group. In this way, one could expect similar qualitative results in the complete case and, in particular, by using the UQG approach. Nevertheless, a full analysis of the corresponding quantum corrected black hole is left for future works.


\section*{Acknowledgements}
R Torres acknowledges the financial support of the Ministerio de Econom\'{\i}a y Competitividad (Spain), projects MTM2014-54855-P.

\appendix

\section{Unimodular spherically symmetric metrics}
Given a spherically symmetric metric written in Schwarzschild-like coordinates as
\begin{equation}\label{nonuni}
ds^2=-f(t,r) dt^2+\frac{dr^2}{f(t,r)} +r^2 d\Omega^2,
\end{equation}
the coordinate change (incidentally, used for the Schwarzschild solution in \cite{Schw})
\begin{equation}\label{coocha}
\rho=\frac{r^3}{3}\ \ ;\ \ x=-\cos \theta
\end{equation}
transforms (\ref{nonuni}) into the unimodular metric (with $det(g_{\mu\nu})=-1$)
\[
ds^2 =-f(t,\rho) dt^2+\frac{d\rho^2}{r(\rho)^4 f(t,\rho)} + r(\rho)^2 \left[\frac{dx^2}{1-x^2}+(1-x^2) d\phi^2 \right].
\]

This metric has coordinate singularities at $x=\pm 1$, which are not relevant from a physical point of view.

The Schwarzschild-de Sitter metric written in Schwarzschild-like coordinates has precisely the form \ref{nonuni} with
\[
f(t,r)=f(r)=1-\frac{2 G_0 m}{r}-\frac{\Lambda}{3} r^2,
\]
so that by using the coordinate change (\ref{coocha}) one directly gets the expression (\ref{USdS}) with $f(\rho)$ as in (\ref{frho}).
Note that the solutions of $f(\rho)=0$ (that define the positions of the horizons in the spacetime) are well-known non-relevant coordinate singularities of the Schwarzschild-de Sitter solution.

\section{Alternative approach for the long distance behaviour}

\begin{figure}
\includegraphics[scale=.6]{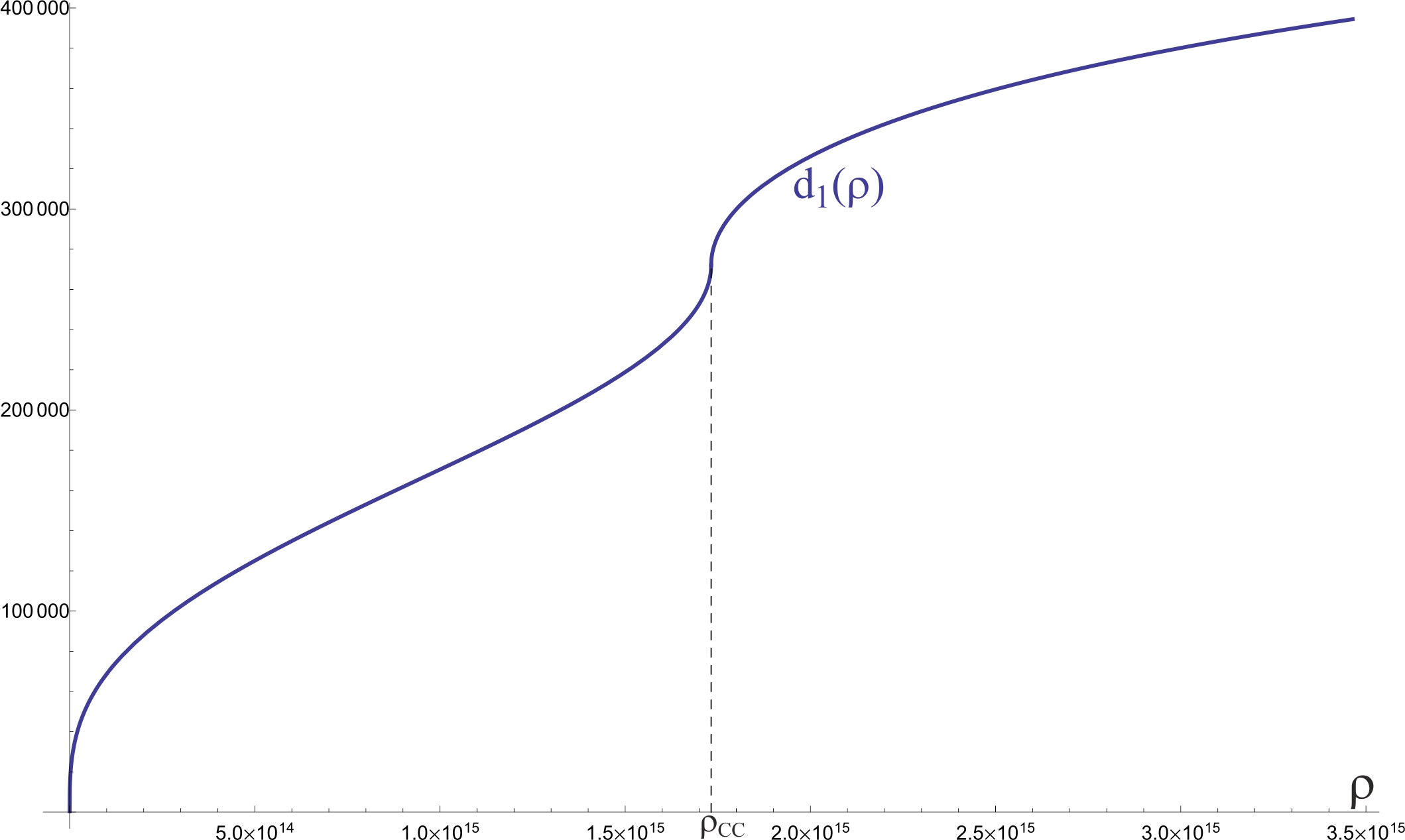}
\caption{\label{figdS} The behaviour of the function $d_1$ in the de Sitter case at both sides of the cosmological horizon $\rho_{CC}$: \textit{arcsin} to its left and \textit{logarithmic} to its right (the part that we are interested in).}
\end{figure}

When considering the long distance behaviour of $d$ for \textit{straight radial curves} one could just consider that at long distances ($\rho\gtrsim \rho_{CC}$) the relevant parameter is $\Lambda$, while the effect of $m$ is negligible. In this way, one could get the approximate behaviour of the distance scale by simply considering the de Sitter spacetime, which can provide us with some analytical results. We will explicit these results in this appendix in order to show that they coincide with ours and to clarify some misunderstandings in the literature. In the de Sitter case we have the metric (\ref{USdS}) with
\[
f(\rho)=1-\frac{\Lambda}{3} (3 \rho)^{2/3},
\]
a unique (cosmological) horizon $\rho_{CC}=\sqrt{3/\Lambda^3}$ and
\[
d_1(\rho)=\int_0^\rho \frac{d\rho}{(3 \rho)^{2/3}\sqrt{|f(\rho)|}}.
\]
If we consider $\rho\leq\rho_{CC}$, we directly get
\[
d_1(\rho\leq\rho_{CC})=\sqrt{\frac{3}{\Lambda}} \arcsin\left(\sqrt{\frac{\Lambda}{3}} \sqrt[3]{3 \rho}\right).
\]
However, for $\rho>\rho_{CC}$ the integral diverges at the horizon and we have to perform it in two steps (from $\rho=0$ to $\rho_{CC}$ and from $\rho_{CC}$ to our desired $\rho$). Thus
\[
d_1(\rho>\rho_{CC})=\frac{\pi}{2}\sqrt{\frac{3}{\Lambda}}+\int_{\rho_{CC}}^\rho \frac{d\rho}{(3 \rho)^{2/3}\sqrt{|f(\rho)|}}.
\]
This admits a (lengthy) analytical expression. Nevertheless, the key here is that $f$ changes its sign at the horizon. In this way, the \textit{arcsin} dependence before the horizon becomes \textit{logarithmic} beyond the horizon and we have
\[
d_1(\rho>\rho_{CC})=\frac{1}{\sqrt{3\Lambda}} \ln(\Lambda^{3/2} \rho)+ \mathcal O (\rho^0),
\]
what is the expected long distance behaviour that we had obtained for the Schwarzschild-de Sitter case (\ref{longd1}). The complete behaviour of $d_1$ in the de Sitter case is shown in figure \ref{figdS}.

\end{document}